\documentclass[Physsubmission, Phys,cite]{SciPost}

\hypersetup{
    colorlinks,
    linkcolor={red!50!black},
    citecolor={blue!50!black},
    urlcolor={blue!80!black}
}

\usepackage{cite}
\usepackage[bitstream-charter]{mathdesign}
\DeclareSymbolFont{usualmathcal}{OMS}{cmsy}{m}{n}
\DeclareSymbolFontAlphabet{\mathcal}{usualmathcal}

\begin{document}

\begin{center}{\Large \textbf{
      Efficient Scheme for
      Time-dependent Thermal Pure Quantum State:
      Application to the Kitaev Model with Armchair Edges
      \\
}}\end{center}

\begin{center}
Hirokazu Taguchi\textsuperscript{1},
Akihisa Koga\textsuperscript{1$\star$} and
Yuta Murakami\textsuperscript{2}
\end{center}

\begin{center}
{\bf 1} Department of Physics, Tokyo Institute of Technology, Meguro, Tokyo 152-8551, Japan
\\
{\bf 2} Center for Emergent Matter Science, RIKEN, Wako 351-0198, Japan
\\
* koga@phys.titech.ac.jp
\end{center}

\begin{center}
\today
\end{center}


\definecolor{palegray}{gray}{0.95}
\begin{center}
\colorbox{palegray}{
  \begin{tabular}{rr}
  \begin{minipage}{0.1\textwidth}
    \includegraphics[width=30mm]{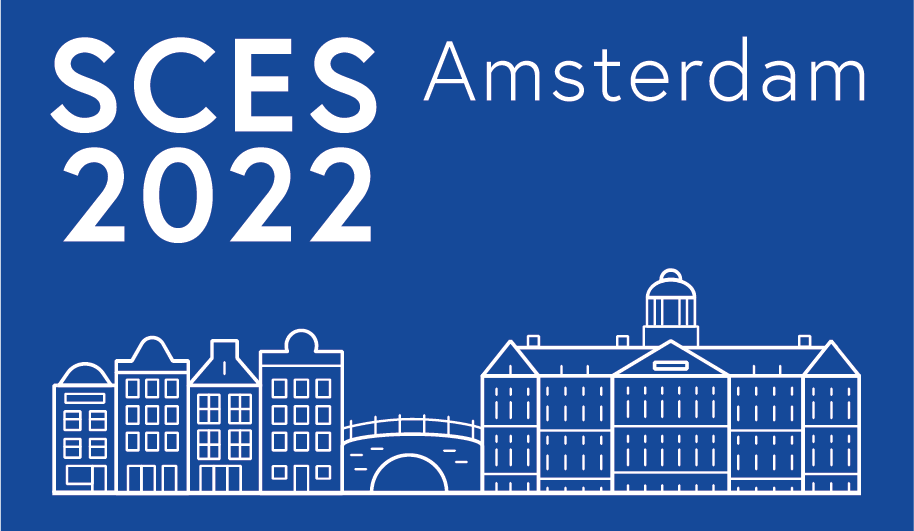}
  \end{minipage}
  &
  \begin{minipage}{0.85\textwidth}
    \begin{center}
    {\it International Conference on Strongly Correlated Electron Systems\\ (SCES 2022)}\\
    {\it Amsterdam, 24-29 July 2022} \\
    \doi{10.21468/SciPostPhysProc.?}\\
    \end{center}
  \end{minipage}
\end{tabular}
}
\end{center}

\section*{Abstract}
{\bf
  We consider the time-dependent thermal pure quantum state method
  and introduce the efficient scheme to evaluate the change in physical quantities
  induced by the time-dependent perturbations,
  which has been proposed in our previous paper
  [H. Taguchi et al., Phys. Rev. B 105, 125137 (2022)].
  Here, we treat the Kitaev model to consider
  the Majorana-mediated spin transport,
  as an example.
  We demonstrate how efficient our scheme is
  to evaluate spin oscillations induced by the magnetic field pulse.
}

\vspace{10pt}
\noindent\rule{\textwidth}{1pt}
\tableofcontents\thispagestyle{fancy}
\noindent\rule{\textwidth}{1pt}
\vspace{10pt}

\section{Introduction}
\label{sec:intro}
Thermal pure quantum (TPQ) state method is
one of the powerful methods to examine
the thermodynamic quantities in the finite clusters~\cite{Sugiura_2012,Sugiura_2013,Endo_2018}
and it successfully has been applied to the quantum spin systems such as
the Heisenberg models on the frustrated Kagome lattices~\cite{Sugiura_2013,Morita_2020,Misawa_2020} and
the Kitaev model~\cite{Yamaji_2016,Tomishige_2018,Nakauchi_2018,KogaS1_2018,Oitmaa_2018,Suzuki_2019,KogaMix_2019}.
Recently, the time-dependent quantities have been examined
in the framework of the TPQ method~\cite{Endo_2018}.
Since the TPQ state is not the eigenstate of the Hamiltonian,
the expectation value, in general, depends on the time $t$
even without the time-dependent Hamiltonian,
yielding to unphysical oscillations.
We have proposed the scheme to reduce ill oscillations
in physical quantities~\cite{Taguchi2}.
In this study, we briefly explain the time-dependent TPQ method
and our scheme.
We then demonstrate an advantage of our scheme,
considering the Majorana-mediated spin transport
in the Kitaev model~\cite{Minakawa_2020,KogaSpin_2020,Taguchi_2021}.

This paper is organized as follows.
In Sec.~\ref{sec2}, we introduce the Kitaev model
with edges and explain the TPQ method.
In Sec.~\ref{sec3}, we demonstrate the numerical results obtained by
the TPQ method
to clarify that our scheme has an advantage 
in evaluating the local physical quantities.
A summary is given in the last section.

\section{Model and Method}\label{sec2}
\subsection{Kitaev model with edges}
In the study, we consider the Kitaev model~\cite{Kitaev_model,Motome_Nasu_review},
which is one of the quantum spin models on the honeycomb lattice
and is composed of the direction dependent Ising interactions.
It is known that there exists a local conserved quantity
defined on each plaquette.
This leads to the important feature in the Kitaev model;
the quantum spin liquid state is realized, where
the magnetic moments and long-range spin-spin correlations are exactly zero. 
\begin{figure}[h]
  \centering
  \vspace{-5mm}
  \includegraphics[width=0.5\linewidth]{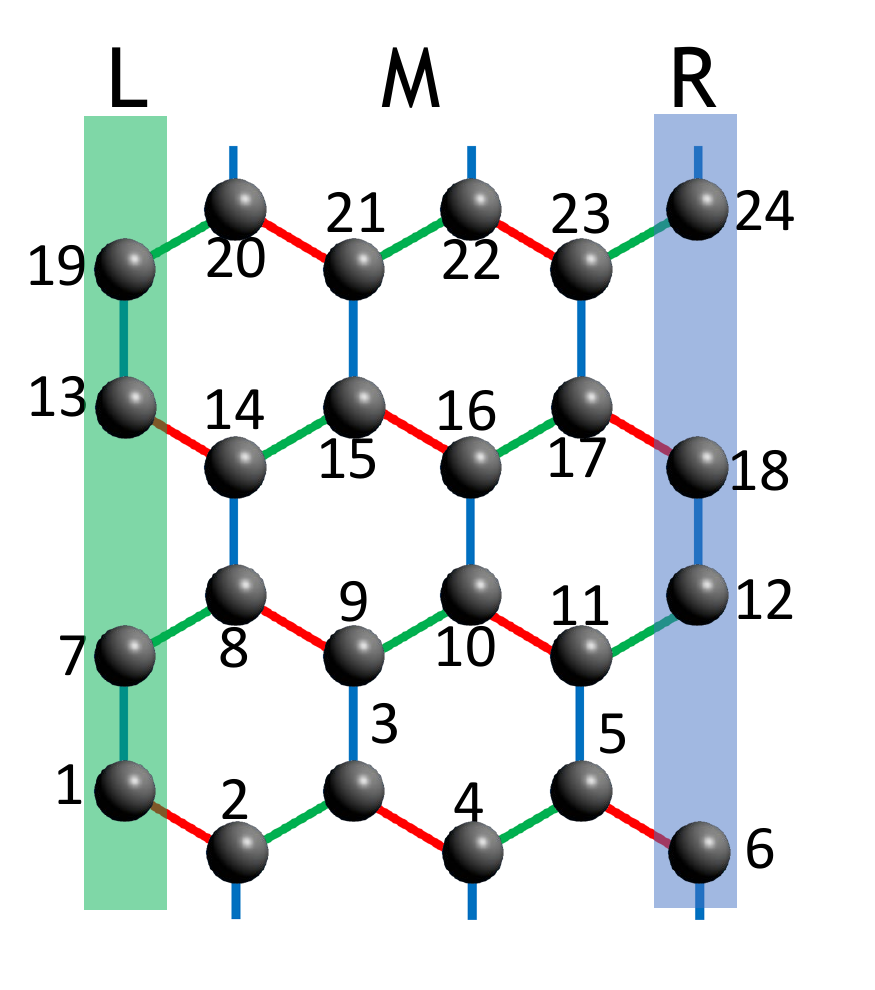}
  \vspace{-5mm}
  \caption{24-site Kitaev cluster with armchair edges.
    Green, red, and blue lines indicate $x$, $y$, and $z$ bonds, respectively.
    The static magnetic field $h_R$ is applied in the right (R) region
    and no magnetic field is applied in the middle (M) region.
    A time-dependent pulsed magnetic field is introduced in the left (L) region.
  }
  \label{fig:system}
\end{figure}
Nevertheless, in the Kitaev model,
the spin excitations propagate,
which are mediated by the itinerant Majorana fermions~\cite{Minakawa_2020}.
This phenomenon can be observed in the Kitaev system excited
by the magnetic field pulse.
One of the simple setups is the system with armchair edges,
as shown in Fig.~\ref{fig:system}. 
The system is composed of the left (L) and right (R) edge regions, and the middle (M) region,
and the tiny static magnetic field is applied only to the R region.
The magnetic pulsed field is introduced to the L region.
The Hamiltonian is given by the static and time-dependent parts as
\begin{eqnarray}
  H(t)&=&H_0+H'(t),\\
  H_0&=&-J\sum_{\langle i,j\rangle_x}S_i^x S_j^x-J\sum_{\langle i,j\rangle_y}S_i^y S_j^y-J\sum_{\langle i,j\rangle_z}S_i^z S_j^z
  -h_R\sum_{i\in R}S_i^z,\\
  H'(t)&=&-h_L(t)\sum_{i\in L}S_i^z,
\end{eqnarray}
where $\langle i, j\rangle_\mu$ indicates the nearest-neighbor pair on the $\mu(=x,y,z)$-bonds.
The $x$-, $y$-, and $z$-bonds are shown as green, red, and blue lines
in Fig.~\ref{fig:system}.
$S_i^\mu$ is the $\mu$ component of an $S = 1/2$ spin operator at the $i$th site and
$J$ is the exchange coupling between the nearest-neighbor spins.
$h_R(=0.01J)$ represents the static magnetic field in the R region.
The time-dependent magnetic field in the L region is given by the Gaussian form as
\begin{align}
  h_L(t) = \frac{A}{\sqrt{2\pi}\sigma}\exp{\left[\frac{t^2}{2\sigma^2} \right]},
  \label{eq:hLt}
\end{align}
where $A$ and $\sigma$ are strength and width of the pulse.
Here, we set $\sigma = 2/J$ and $A = 1$.
In this study, we examine the expectation value of the local quantity
after the magnetic pulse is introduced in the L region.

\subsection{Thermal Pure Quantum state method}

Here, we explain the TPQ method to examine the expectation value of the
physical quantity at finite temperatures.
When $t\rightarrow -\infty$, the system is described by the static Hamiltonian
$H_0$ and the expectation value for a certain operator $\hat{O}$
is given by the trace calculations as
\begin{align}
  \langle\hat{O}\rangle = \frac{1}{Z_0}\text{Tr}\left[ \hat{O} e^{- \beta H_0} \right],
\end{align}
where $\beta=1/T$, $T$ is the temperature,
$Z_0(=\text{Tr}\left[ e^{- \beta H_0} \right])$ is the partition function.
It is known that at zero temperature ($T=0$),
only the ground state contributes to the expectation value.
On the other hand, at finite temperatures,
all eigenstates need to evaluate the expectation value,
which make it hard to treat larger clusters numerically.
Instead, we use the TPQ state method~\cite{Sugiura_2012, Sugiura_2013}.
The expectation value is represented as
\begin{align}
  \langle \hat{O}\rangle = \langle \Psi_T| \hat{O}|\Psi_T\rangle,
\end{align}
where $|\Psi_T\rangle$ is the TPQ state at the temperature $T$.
In contrast to the former method, one does not have to calculate all eigenvalues and eigenstates,
and thereby the TPQ state method has an advantage in treating larger systems.

We briefly describe how to construct the TPQ state.
A TPQ state at $T\rightarrow\infty$ is simply given by a random vector,
\begin{eqnarray}
  |\Psi_0\rangle = \sum c_i |i\rangle,~\label{random}
  \end{eqnarray}
where $\{ c_i\}$ is a set of random complex numbers satisfying $\sum_i |c_i|^2=1$
and $|i\rangle$ is an arbitrary Hilbert basis.
By multiplying a certain TPQ state by the Hamiltonian,
the TPQ states at lower temperatures are constructed.
The $k$th TPQ state is represented as
\begin{eqnarray}
  |\Psi_k\rangle = \frac{(L-H_0)|\Psi_{k-1}\rangle}{||(L-H_0)|\Psi_{k-1}\rangle||},
\end{eqnarray}
where $L$ is a constant value, which is larger than the maximum
eigenvalue of the Hamiltonian $H_0$.
The corresponding temperature is given by
\begin{eqnarray}
  T_k=\frac{L-E_k}{2k},
\end{eqnarray}
where $E_k(=\langle\Psi_k|H_0|\Psi_k\rangle)$ is the internal energy.
The thermodynamic quantities such as entropy and specific heat
can be obtained from the internal energy and temperature.
We repeat this procedure until $T_k=T$ and
obtain the TPQ state at the temperature $T$, $|\Psi_T\rangle$.

The time-dependent quantities are also evaluated in the framework of the TPQ method~\cite{Endo_2018}.
The expectation value at time $t$ for an operator $\hat{O}$ is given as
\begin{eqnarray}
  \langle\hat{O}(t)\rangle
  &=&\frac{1}{Z_0}{\rm Tr}\left[\hat{O}(t)e^{-\beta H_0}\right], \notag\\
  &=& \langle\Psi_T| \hat{O}(t)|\Psi_T\rangle, \notag \\
  &=&\langle \Psi_T(t)| \hat{O} |\Psi_T(t)\rangle,
\end{eqnarray}
where $\hat{O}(t)=U^\dag(t) \hat{O} U(t)$, $|\Psi_T(t)\rangle=U(t)|\Psi_T\rangle$,
and $U(t)$ is the time evolution operator.
Therefore, we can discuss the time-evolution of the system
in terms of the time-evolution of the TPQ state.
When one discusses the real-time dynamics triggered by the Hamiltonian $H'(t)$,
it is useful to examine a change in the quantities as,
\begin{eqnarray}
  \Delta O(t)=\langle \hat{O}(t)\rangle-\langle \hat{O} \rangle.\label{Delta}
\end{eqnarray}
In the following, we focus on this quantity.

The TPQ method has an advantage in treating larger clusters, while
we sometimes suffer from unavoidable numerical problems.
When the TPQ method is applied to the finite cluster,
the obtained results are sensitive to its size and/or shape.
This is due to, at least, two effects.
One of them is that
low energy properties in the thermodynamic limit cannot be described correctly
in terms of finite clusters.
Therefore, the large system size dependence
in the physical quantities appears at low temperatures
although the TPQ method reproduces the correct results at higher temperatures.
The other is the random dependence in the initial TPQ state.
In general, this can be excluded, by taking a statistical average of
the results for independent TPQ states.
Nevertheless, we sometimes meet with difficulty in evaluating time-dependent quantities
due to their large variance.
This originates from the fact that
each TPQ state is not an eigenstate of the Hamiltonian,
leading to ill oscillations in the physical quantities
with respect to time even without time-dependent perturbations,
unless the quantities are conserved ones.
Namely, the sample dependence is somewhat large even at high temperatures.

To avoid the latter problem, we prepare two time-dependent TPQ states from
the common TPQ state as,
$|\Psi_T(t)\rangle$ and $|\Psi_T^0(t)\rangle=U_0(t)|\Psi_T\rangle$~\cite{Taguchi2},
where $U_0(t)$ is the time-evolution operator for the system described by $H_0$.
Then, we calculate $\langle\langle \hat{O}(t)\rangle\rangle=\langle \Psi_T^0(t)|\hat{O}|\Psi_T^0(t)\rangle$ instead of $\langle \hat{O} \rangle$
and evaluate the change in the quantities (\ref{Delta}),
where unphysical oscillations should be canceled.
This allows us to obtain $\Delta O(t)$ efficiently and
to discuss correctly how the time-dependent Hamiltonian $H'(t)$ affects the system
at finite temperatures.

\section{Results}\label{sec3}
We here demonstrate how efficient our scheme is to evaluate
the change in the quantities at finite temperatures.
To clarify this, 
we treat the 24-site Kitaev cluster with armchair edges (Fig.~\ref{fig:system})
and evaluate the change in the moment at the $i$th site $\Delta S_i^z(t)$
by means of the TPQ states constructed from a certain random state. 
The quantity is described in two distinct ways as
\begin{eqnarray}
  \langle S_i^z(t)\rangle - \langle S_i^z\rangle &=& \langle\Psi_T(t) | S_i^z | \Psi_T(t)\rangle
  - \langle\Psi_T| S_i^z | \Psi_T\rangle, \label{eq.1} \\
  \langle S_i^z(t)\rangle - \langle\langle{S_{i}^z(t)}\rangle\rangle &=& \langle{\Psi_T(t) | S_i^z | \Psi_T(t)}\rangle  - \langle{\Psi_T^0(t) | S_i^z | \Psi_T^0(t)}\rangle. \label{eq.3}
\end{eqnarray}
Then, we compare the results of the conventional scheme eq.~(\ref{eq.1}) and our scheme
eq.~(\ref{eq.3}).

We calculate the change in the spin moments in the system
after the magnetic pulse is introduced in the L region.
In the M region, the moments are never induced due to
the existence of the local conversed quantities~\cite{Kitaev_model}.
In fact, we have confirmed the absence of the moments
in the framework of the TPQ method (not shown).
Now, we focus on two sites ($i=12$ and $24$) in the R region.
These sites are equivalent
since the Kitaev cluster treated here has a translational symmetry
in the direction along the edge, as shown in Fig.~\ref{fig:system}.
\begin{figure}[htb]
  \centering
  \includegraphics[width=0.7\linewidth]{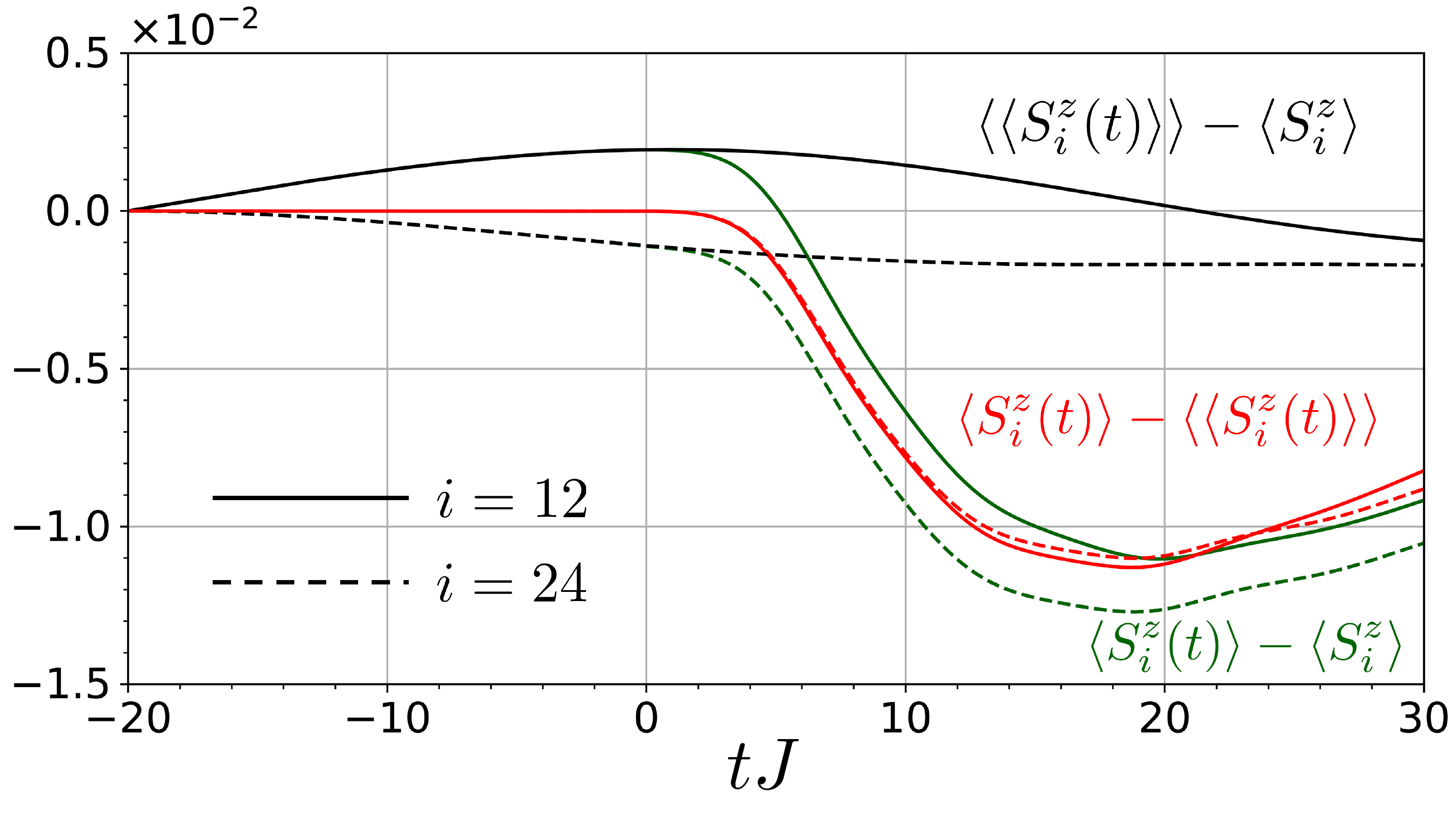}
  \caption{Green, red, and black lines indicate
    time-dependent quantities at $T/J=0.05$,
    which are calculated by the formulations eqs.~(\ref{eq.1}), (\ref{eq.3}),
    and (\ref{eq.2}), respectively.
  }
  \label{fig:Demo}
\end{figure}
Figure~\ref{fig:Demo} shows the change in the moment in the R region
at the temperature $T/J=0.05$,
which is obtained from only one TPQ state $|\Psi_T\rangle$.
We start with the simulation from the time $t=-20/J$,
where $h_L(t)$ is small enough.
Nevertheless, we find that
the changes in the spin moment described by eq.~(\ref{eq.1})
are immediately induced in both sites $i=12$ and $24$
although the latter may be invisible.
To understand this phenomenon, we also consider the quantity
\begin{eqnarray}
  \langle\langle S^z_i(t)\rangle\rangle-\langle S^z_i\rangle&=&
  \langle\Psi^0_T(t) | S_i^z | \Psi^0_T(t)\rangle
  - \langle\Psi_T| S_i^z | \Psi_T\rangle,\label{eq.2}
\end{eqnarray}
which is calculated in terms of the time-evolution operator $U_0(t)$.
We find in Fig.~\ref{fig:Demo} that, around $t\sim -20/J$,
the quantities are the same as the results obtained from eq.~(\ref{eq.1}).
This means the existence of unphysical oscillations
originating from the initial TPQ state.
Beyond $t\sim 2/J$, we find that 
the difference in the results obtained
from eqs.~(\ref{eq.1}) and (\ref{eq.2})
becomes larger.
This suggests that the physically meaningful oscillations are induced by
the time-dependent perturbations (magnetic pulse introduced in the L region).
In fact, by taking the statistical average in these quantities
obtained from many independent TPQ states,
we can confirm that the spin oscillations are correctly described
by eq.~(\ref{eq.1}).
However, for the result obtained by one TPQ state,
the induced oscillation around $t\sim 4/J$ and
the initial unphysical oscillation
are of the same order in this case.
Therefore, the statistical average for the results obtained
from a large number of TPQ states
may be necessary to obtain the numerically reliable results.
On the other hand, we clearly find that the formulation eq.~(\ref{eq.3})
correctly describes the change in the moment;
no oscillation behavior appears before the Gaussian pulse is introduced in the L region, and
the quantities for both sites are almost the same,
which is consistent with the symmetry argument in the system.
It is naively expected that the accurate results should be obtained
by the statistical average for a smaller number of samples.
Therefore, we can say that our scheme (\ref{eq.3}) has an advantage
in evaluating the local physical quantities.

\section{Conclusion}
We have treated the Kitaev model with edges
to examine the time-evolution of the spin moments
by means of the time-dependent thermal pure quantum state method.
We have explained the detail of our scheme proposed in our previous paper,
where two kinds of the time-evolution operators are applied 
to the common TPQ state.
Then, we have demonstrated that ill oscillations in the physical quantities, 
which originate from the fact that each TPQ state is not an eigenstate and 
causes a problem in the conventional time-evolution scheme, are suppressed.
We have evaluated the change in the moment at each site in the Kitaev model.

\section*{Acknowledgements}
Parts of the numerical calculations are performed
in the supercomputing systems in ISSP, the University of Tokyo.


\paragraph{Funding information}
This work was supported by Grant-in-Aid for Scientific Research from
JSPS, KAKENHI Grant Nos.
JP17K05536, JP19H05821, JP21H01025, JP22K03525 (A.K.),
JP20K14412, JP21H05017 (Y.M.), and JST CREST Grant No. JP-MJCR1901 (Y.M.). 

\bibliography{refs.bib}

\nolinenumbers

\end{document}